\documentclass[twocolumn,epsf,prl,aps]{revtex4}
\usepackage{amssymb}
\usepackage{amsmath}
\usepackage{amsfonts}
\usepackage{amscd}
\usepackage{graphics}
\usepackage{epic}
\usepackage{eepic}
\usepackage{color}
\usepackage{epsfig}
\usepackage{latexsym}
\usepackage{graphicx}

\begin{document}

\def\ra{\rangle}
\def\la{\langle}
\def\bege{\begin{equation}}
\def\ende{\end{equation}}
\def\begarr{\begin{eqnarray}}
\def\endarr{\end{eqnarray}}
\def\ha{{\hat a}}
\def\hb{{\hat b}}
\def\hu{{\hat u}}
\def\hv{{\hat v}}
\def\hc{{\hat c}}
\def\hd{{\hat d}}
\def\hi{\hangindent=45pt}
\def\v{\vskip 12pt}

\def\abs#1{ \left| #1 \right| }
\def\lg#1{ | #1 \rangle }
\def\rg#1{ \langle #1 | }
\def\lrg#1#2#3{ \langle #1 | #2 | #3 \rangle }

\newcommand{\bra}[1]{\left\langle #1 \right\vert}
\newcommand{\ket}[1]{\left\vert #1 \right\rangle}
\newcommand{\bx}{\begin{matrix}}
\newcommand{\ex}{\end{matrix}}
\newcommand{\be}{\begin{eqnarray}}
\newcommand{\ee}{\end{eqnarray}}
\newcommand{\nn}{\nonumber \\}
\newcommand{\no}{\nonumber}
\newcommand{\de}{\delta}
\newcommand{\lt}{\left\{}
\newcommand{\rt}{\right\}}
\newcommand{\lx}{\left(}
\newcommand{\rx}{\right)}
\newcommand{\lz}{\left[}
\newcommand{\rz}{\right]}
\newcommand{\inx}{\int d^4 x}
\newcommand{\n}{\nonumber}
\newcommand{\pu}{\partial_{\mu}}
\newcommand{\pv}{\partial_{\nu}}
\newcommand{\au}{A_{\mu}}
\newcommand{\av}{A_{\nu}}
\newcommand{\p}{\partial}
\newcommand{\ts}{\times}
\newcommand{\ld}{\lambda}
\newcommand{\al}{\alpha}
\newcommand{\bt}{\beta}
\newcommand{\ga}{\gamma}
\newcommand{\si}{\sigma}
\newcommand{\ep}{\varepsilon}
\newcommand{\vp}{\varphi}
\newcommand{\ro}{\rho}
\newcommand{\tu}{\tau}
\newcommand{\dg}{\dagger}

\title{Sub-Shot-Noise Quantum Optical Interferometry:\\
A Comparison of Entangled State Performance within a Unified
Measurement Scheme}

\author{Yang Gao}\email{ygao1@lsu.edu}
\author{Hwang Lee}\email{hwlee@phys.lsu.edu}

\affiliation{
Hearne Institute for Theoretical Physics,
Department of Physics and Astronomy, 
Louisiana State University, Baton Rouge, LA 70803}

\date{\today}


\begin{abstract}
Phase measurement using a lossless Mach-Zehnder
interferometer with certain entangled $N$-photon states can
lead to a phase sensitivity of the order of $1/N$, the Heisenberg
limit. However, previously considered output measurement schemes are different for
different input states to achieve this limit. 
We show that it is possible to achieve this limit just by the
parity measurement for all the commonly proposed entangled states.
Based on the parity measurement scheme, the reductions of the phase 
sensitivity in the presence of photon loss are examined for the various input states.

\end{abstract}

\maketitle

The notion of quantum entanglement holds great promise for certain 
computational and communication tasks.
It is also at the heart of metrology
and precision measurements in extending their capabilities
beyond the so-called standard quantum limit \cite{caves81,giovannetti06,higgins07}. 
For example, the phase sensitivity of a usual two-port interferometer has 
a shot-noise limit (SL) that scales as $1/\sqrt{N}$, where $N$ is the
number of the photons entering the input port.
However, 
a properly correlated
Fock-state input for the Mach-Zehnder interferometer can
lead to an improved phase sensitivity that scales as $1/N$, i.e.,
the Heisenberg limit (HL) \cite{yurke86,yuen86,ou96,dowling98}. 
In the subsequent development, the dual Fock-state \cite{holland93} 
and the so-called intelligent
state \cite{hillery93,brif96} 
were proposed to
reach a sub-shot-noise sensitivity as well. 
Recently, 
much attention has been paid to the so-called NOON
state to reach the exact HL in interferometry 
as well as super-resolution imaging \cite{lee02,walther04,resch07,nagata07}.

The utilization of those quantum correlated input states
are accompanied by various output measurement schemes.
In some cases the conventional measurement scheme of photon-number
difference is used, whereas a certain 
probability distribution \cite{hradil95,sanders95,kim98,pooser04},
a specific adaptive measurement \cite{wiseman95,berry00,armen02},
and the parity measurement
are used for other cases.

Gerry and Campos first showed the use of the parity measurement for
the ``maximally entangled state''--the NOON state--of light 
to reach the exact HL \cite{gerry01}, 
following the earlier suggestion of the HL spectroscopy
with $N$ two-level atoms \cite{bollinger96}.
Campos, Gerry, and Benmoussa later suggested that 
the parity measurement scheme can also be used for 
the dual Fock state inputs by
comparing the quantum state {\em inside} the interferometer
with the NOON state \cite{campos03}.
In this paper we show that the
parity measurement can
be used as a detection scheme for sub-shot-noise interferometry
with the correlated Fock states first proposed 
by Yurke, McCall, and Klauder \cite{yurke86},
as well as with
the intelligent states first suggested
by Hillery and Mlodinow \cite{hillery93}.
Extension of its use for all these input states
then promote the parity measurement to a kind of 
universal detection scheme for 
quantum interferometry.
Then, based on such a universal detection scheme
comparisons of performance of various
quantum states can be made in a common ground.
As an example,
we present a comparison of the phase sensitivity reduction for 
various quantum states of light in the presence of photon loss.

In order to describe the notations, 
we briefly review the group theoretical formalism of 
Mach-Zehnder interferometer. 
The key point of such a formalism is that 
any passive lossless four-port optical system 
can be described by the SU(2) group \cite{yurke86}. 
First, we use the mode annihilation operators $a_{in(out)}$
and $b_{in(out)}$, which satisfy boson commutation relations, to
represent the two light beams entering (leaving) the beam splitter
(BS), respectively. 
Then the action of BS takes the form 
\be \lx \bx a_{out} \\
b_{out} \ex \rx =\lx \bx
e^{i (\al+\ga)/2}\cos\frac{\bt}{2}  & 
 e^{-i (\al-\ga)/2}\sin\frac{\bt}{2} \\
- e^{i(\al-\ga)/2}\sin\frac{\bt}{2} & e^{-i (\al+\ga)/2}
\cos\frac{\bt}{2} \ex \rx \lx \bx a_{in} \\ b_{in} \ex \rx .\ee
Here $\al$, $\bt$, and $\ga$ denote the Euler angles
parameterizing SU(2), and they are
related to the complex transmission and reflection coefficients. 
Through the Schwinger representation of angular momentum
we can construct the operators for the angular momentum
and for the occupation number
from the mode operators $a$ and $b$, 
\be \bf{J}=\lx \bx J_{x} \\
J_{y} \\ J_{z} \ex \rx &=& \frac{1}{2} \lx \bx a b^\dg + b a^{\dg} \\
i(a b^\dg-b a^\dg) \\ a a^\dg - b b^\dg \ex \rx, \ee 
and $ N= a^\dg a + b^\dg b$. 
The commutation relations $[a,b]=[a,b^\dg]=0$
and $[a,a^\dg]=[b,b^\dg]=1$ lead to the relation 
${\bf J} \times {\bf J} = i {\bf J}$.
The Casimir invariant has the form
$J^2=J_x^2+J_y^2+J_z^2=(N/2)(N/2+1)$ that commutes with $J_i$ and $N$. 
Next, it was shown that the operation of the BS
is equivalent to \cite{yurke86} 
\be 
\mathbf{J}_{out} =e^{i\al J_z}e^{i\bt
J_y}e^{i\ga J_z}\mathbf{J}_{in} e^{-i\ga J_z}e^{-i\bt J_y}e^{-i\al
J_z},
\ee 
in the Heisenberg picture, and to 
\be \lg { \text {out}} =
e^{-i\al J_z}e^{-i\bt J_y}e^{-i\ga J_z} \lg { \text {in} }, \ee in
the Schr\"odinger picture. 
If we use the symbols $j$ and $m$ to
indicate the eigenvalues of $N/2$ and $J_z$, then the theory of
angular momentum tells that the representation Hilbert space is
spanned by the complete orthonormal basis $\lg{j,m}$ with
$m\in[-j,j]$, which can also be labeled by the Fock states of the
two modes, $\lg {j,m}=\lg {j+m}_a \lg{j-m}_b$. 
In terms of this language, 
we may make the geometrical interpretation of the elements
of a Mach-Zehnder interferometer. For example, the effect of a
50/50 BS, which leads a $\pm \pi/2$ rotation around the $x$ axis 
(given by the unitary transformation $e^{\pm i(\pi/2)J_x}$),
is
equivalent to the transformation
\be \lx \bx a_{out} \\
b_{out} \ex \rx = \frac{1}{\sqrt 2}\lx \bx 1 & \mp i \\
\mp i & 1 \ex \rx \lx \bx a_{in} \\
b_{in} \ex \rx .\ee  
Similarly, the relative phase shift $\vp$ acquired between the
two arms of the Mach-Zehnder interferometer can be
expressed by 
$a_{out} = a_{in}$, 
$b_{out} = e^{i \vp} b_{in}$,
or by the unitary transformation
$e^{-i\vp J_z}$ equivalently.
\begin{figure}[t]
\centerline{\includegraphics[scale=0.45]{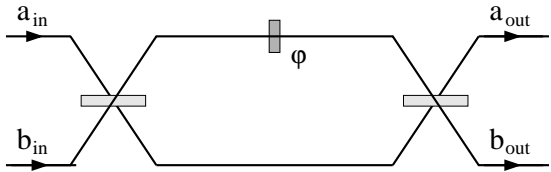}}
\caption{\label{mzi} Schematic of the Mach-Zehnder interferometer.
The angle $\varphi$ denotes the relative phase difference between
the arms.
Note that the Yurke, dual-Fock, and intelligent states are inserted
to the left of the first beam splitter, and NOON to the right.}
\end{figure}
The Mach-Zehnder interferometer can be illustrated schematically in FIG.~1, 
where the two light beams $a$ and $b$ first enter the $\rm BS_+$, 
and then acquires a relative phase shift $\vp$, and finally
pass through the $\rm BS_-$. 
The photons leaving the $\rm BS_-$ are
counted by detectors $\rm{D_a}$ and $\rm{D_b}$. 
Therefore, in the language of the group theory, 
the input states of $\rm BS_+$ and
the output states of $\rm BS_-$ is connected 
by a simple unitary
transformation
$U = e^{i(\pi/2) J_x}e^{-i\vp J_z}e^{-i(\pi/2)
J_x}=e^{-i\vp J_y}$ \cite{bs}.

The information on the phase
shift $\vp$ is inferred from the photon statistics of the output beams. 
There are many statistical methods to extract such information. 
The most common one is to use the difference between
the number of photons in the two output modes,
$N_d=a_{out}^\dagger a_{out}-b_{out}^\dagger b_{out}$, or
equivalently, $J_{z,out}=N_d/2$. 
The minimum detectable
phase shift then can be estimated by \cite{scully97}
\be 
\de \vp = \frac{\Delta J_{z,out}}{|\p \langle J_{z,out} \rangle /\p \vp|}, \label{phj}
\ee 
where $ \Delta J_{z,out}=\sqrt{\langle J_{z,out}^2
\rangle-\langle J_{z,out} \rangle^2}$. 
The expectation value of
$J_{z,out}$ and $J_{z,out}^2$ are calculated by  
$\langle J_{z,out} \rangle 
= \langle {\text{in}} |{J_{z,out}} |{\text{in}} \rangle 
= \langle {\text{in}} |{U^\dg J_{z,in} U} |{\text{in}} \rangle
$,
$
\langle
J_{z,out}^2 \rangle 
= \langle {\text{in}}| {J_{z,out}^2} |{\text{in}} \rangle 
= \langle {\text{in}} |{U^\dg J_{z,in}^2 U} |{\text{in}}\rangle
$, 
and 
$U^\dg J_{z,in}^n U = (-\sin \vp J_{x,in} +\cos \vp J_{z,in} )^n
$.

Now the application of the group formalism to analyze the phase
sensitivity of the ideal Mach-Zehnder interferometer is
straightforward. 
Let us first consider the correlated photon-number 
states \cite{yurke86,yuen86,dowling98}.
In particular, the so-called Yurke state has the form 
$\lg {\text{in}} = \lz\lg {j,0}+\lg{j,1}\rz/\sqrt 2$, which is one of the 
earliest proposals of utilizing 
the correlated photon-number states \cite{yurke86}).
A simple calculation for the Yurke-state input gives 
\be
\de \vp = \frac{\lt\lz j(j+1)-1 \rz \sin^2 \vp  + \cos^2 \vp \rt ^{1/2}}{
|{\sqrt {j(j+1)} \cos \vp + \sin \vp}| } , 
\ee 
which has its minimum value 
$\de \vp_{min} \approx {1 / \sqrt{j(j+1)}}$
when $\sin \vp \approx 0$. 
Hence, when the Yurke state is fed into
the input ports of an interferometer, 
the minimum of $\de \vp$ has
the order of $2/N$ limit since $j=N/2$. 
We should bear in mind that
the minimum phase sensitivity is achieved only at particular
values of $\vp\approx 0$. For other values of $\vp$ the phase
sensitivity is decreased. 
However, one can always control the phase
shift by a feed-back loop which keeps $\vp$ at any particular
value.

On the other hand, the parity measurement, represented by
the observable
$P=(-1)^{b^\dg b}=e^{i \pi (j-J_z)}$ has an advantage 
when the simple photon number counting method ceases 
to be appropriate to infer the phase shift 
and provides a wider applicability than $J_{z}$.
The parity measurement scheme was first introduced by 
Bollinger, Itano, Wineland, and Heinzen for spectroscopy with
trapped ions of maximally entangled form \cite{bollinger96}.
Gerry and Campos adopted such a measurement scheme 
to the optical interferometry with the NOON state \cite{gerry01}.
The NOON state can be formally written as 
$\lg {\text {NOON}} = [\lg {j,j}+\lg{j,-j}]/\sqrt{2}$.
Note that the NOON state is  {\em not} the input state of MZI, but
the state {\em after} the first beam splitter $\rm BS_+$. 
Hence the output state is described as 
$ 
\lg {\text {out}} = e^{i
(\pi/2)J_x}e^{-i \vp J_z} \lg {\text {NOON}}$. 

The expectation value for the parity operator is then given by 
$\langle P \rangle = 
= i ^ N \rg {\text {NOON}} e^{i \vp J_z} e^{i \pi J_y} 
e^{-i \vp J_z} \lg {\text {NOON}} 
= i^N \lz e^{i N \vp}+ (-1)^N
e^{-i N \vp}\rz /2$, 
so that we have 
\be
\langle P \rangle &=& 
\bigg \{ \bx i^{N+1} \sin N \vp , & \;\;\ N \;\ \text{odd}, \\
i^N \cos N \vp , & \;\;\ N \;\ \text{even}, \ex  \label{noon} \ee
where the identity 
$e^{-i (\pi/2) J_x} e^{-i \pi J_z} e^{i
(\pi/2)J_x}=e^{i \pi J_y}$
is applied. 
Since $P^2=1$, the equation
(\ref{noon}) then immediately leads to the result $\de \vp = 1/N$, exactly. 

Now, let us consider the dual Fock-state as the input state, 
$\lg {j,0}=\lg j_a \lg j_b$. 
Here, if we still use $J_{z,out}$ as our
observable, we have 
$\langle J_{z,out} \rangle = \rg {j,0}
-\sin \vp J_x + \cos \vp J_z \lg {j,0}=0. $
The
expectation value of the difference of the output photon number is now
independent of the phase shift. 
Therefore, in this case the
measurement of $J_{z,out}$ contains no information about the phase shift.
A method of reconstruction of the probability distribution 
has been proposed to avoid this phase independence and to reach 
the Heisenberg limit \cite{holland93,kim98,pooser04}.
More recently, Campos, Gerry, and Benmoussa suggested the use of 
the parity measurement for the dual Fock-state inputs \cite{campos03}.

The expectation value of $P$ can be
derived from $\langle P_{out} \rangle = \rg {\text{in}} e^{i
\vp J_y} P_{in} e^{-i \vp J_y} \lg {\text{in}}$ and 
$\langle
P_{out}^2 \rangle = \rg {\text{in}}  \text{in} \rangle=1$. 
For the dual Fock-state,
we have $\langle P_{out} \rangle ^{\text {d-Fock}} = \rg {j,0}
e^{i \vp J_y} (-1)^{j-J_z} e^{-i \vp J_y} \lg {j,0} = (-1)^j
\,\ d^{j}_{0,0}(2\vp)$, 
where
$d^{j}_{m,n}$ denotes the rotation matrix element: 
$e^{-i\vp
J_y}\lg {j,n} = \sum_{m=-j}^{j}d^{j}_{m,n}(\vp)\lg {j,m}$, and
\be
&& d^{j}_{m,n}(\vp) = (-1)^{m-n} {2^{-m}
}\sqrt{\frac{(j-m)!(j+m)!}{(j-n)!(j+n)!}}
 \nn 
&& \ts \,\ P_{j-m}^{(m-n,m+n)}(\cos\vp)\,\ (1-\cos\vp)^{m-n \over 2}
(1+\cos\vp)^{m+n \over 2}_, \nonumber \ee 
where
$P_{n}^{(\al,\bt)}(x)$ represents the Jacobi polynomial. 
Thus the phase sensitivity is obtained as
$\de \vp^{\text {d-Fock}}= \{ 1- [d^{j}_{0,0}(2\vp) ]^{2}\}^{1/2}
/| {\p d^{j}_{0,0}(2\vp)/\p \vp}|$
for the dual Fock-state, and 
in the limit of $ \vp \to 0$, 
we have 
$\de
\vp ^{\text {d-Fock}} \to 1/\sqrt{2j(j+1)} \sim \sqrt{2}/N$.

If we use the parity measurement scheme for the Yurke-state input, 
we obtain
\be && \langle P_{out} \rangle
^{\text {Yurke}} =  \rg {\text{in}} e^{i \vp J_y} (-1)^{j-J_z}
e^{-i \vp J_y} \lg {\text{in}} \nn & & =
\sum_{m=-j}^{j}\frac{(-1)^{j-m}}{2}\lx {d^{j\ast}_{m,0}}+{
d^{j\ast}_{m,1}}\rx \lx d^{j}_{m,0}+d^{j}_{m,1} \rx \nn & &=
\frac{(-1)^j}{2} \lz d^{j}_{0,0}+d^{j}_{0,1}
-d^{j}_{1,0}-d^{j}_{1,1} \rz (2\vp), \label{dmn}
\ee 
where have
used the following properties of the matrix element \cite{ang} 
in the last line of (\ref{dmn}): 
\be {d^{j\ast}_{m,n}}=d^{j}_{m,n}
=(-1)^{m-n}d^{j}_{n,m}=d^{j}_{-n,-m} \nn \sum_{m=-j}^{j}
{d^{j}_{k,m}(\vp_1)}\,\ d^{j}_{m,n}(\vp_2)=
d^{j}_{k,n}(\vp_1+\vp_2). \ee 
Again, using 
$\de \vp =
\sqrt{1-[\langle P_{out} \rangle ^{\text {Yurke}}]^2
}/{\abs{ \p \langle P_{out} \rangle ^{\text {Yurke}} /\p \vp }}$,
we have $\de
\vp ^{\text {Yurke}} \to 1/\sqrt{j(j+1)} \sim 2/N$, in the limit of $ \vp \to 0$. 
This shows that, for the Yurke state,
the parity measurement scheme leads to the same phase sensitivity
as the $J_{z,out}$ measurement scheme.
The dual-Fock state then performs better than the Yurke-state
by a factor of $\sqrt{2}$ within
the parity measurement scheme.

We can also use parity observable for the intelligent state
entering the first beam splitter $\rm BS_+$ in FIG.~1. 
The intelligent state is defined as the
solution of the equation 
$
\lx J_y+i\eta J_z \rx
\lg {j,m_0,\eta}=\bt \lg {j,m_0,\eta}
$, 
where $\eta^2=(\Delta J_y)^2/(\Delta J_z)^2 $ 
and $m_0$ is an integer belonging to
$[-j,j]$ \cite{hillery93}.
The eigenvalue corresponding to $\lg {j,m_0,\eta}$ is
$\bt=i \!\ m_0 \sqrt{\eta^2-1}$ and the eigenvector 
$\lg {j,m_0,\eta}=\sum_{k=-j}^{j}C_k \lg {j,k}$, 
where an explicit form of the expansion coefficient
$C_k$ is given in Ref.~\cite{brif96}. 
The expectation value of the parity operator is then obtained as 
$\langle
P_{out}\rangle^{\text {Int}} = (-1)^{j} \sum_{k,n=-j}^{j}
C^*_k C_n (-1)^k d^j_{k,n}(2\vp)$. 
It follows that 
from the explicit form of $C_k$'s
the phase sensitivity
scales better with a larger $\eta$ and a smaller $\abs{m_0}$
As
$\eta \to \infty$, the phase sensitivity becomes 
\be 
\de \vp^{\text {Int}} \to
{1 \over \sqrt{2(j^2-m_0^2+j)}} 
\sim {\sqrt{2} \over N}. 
\ee 
On the other hand, as $\eta \to 1$, 
we have $ \de \vp^{\text {Int}} \to {1 / \sqrt{2j}} \sim 1/\sqrt{N}$, 
which is the standard shot-noise limit. 
So the minimum value of $\de \vp$
is only accessible for $m_0=0$. 
This limiting behavior is the same
as the phase sensitivity with $J_z$ measurement at $\vp=0$
\cite{brif96}.
We note that, within the parity measurement scheme,
of all states considered here only the NOON state
reaches exactly the HL \cite{durkin07}.

Now that we have seen we can adopt the parity measurement
as a universal detection scheme for all the commonly used
entangled states, we will use it as a common ground to compare
the effect of photon loss on phase sensitivity, thus we can
put each input state on the same footing. 

The effect of photon loss has been recently studied 
for the NOON states.
Gilbert and coworkers applied a model for loss as
a series of beam splitters in the propagation paths \cite{gilbert06}.
Rubin and Kaushik applied a single beam-splitter model for loss
on the detection operator \cite{rubin07}.
Whereas the two approaches are equivalent, we adopt
the one given in Ref.~\cite{gilbert06} by putting the 
the effect of photon loss in the following form \cite{loudon00}:
$ 
a_{out} = e^{(-i\eta_a \omega/c -K_a/2)L_a}
a_{in}  
+ i \sqrt{K_a} \int_0^{L_a}dz \,\ e^{(-i\eta_a
\omega/c -K_a/2)(L_a-z)} d(z)$,
where $\eta_i$ is the index of
refraction for arm $i$ of the interferometer, $K_i$ is the
absorption coefficient, and $L_i$ is the path length. 
The annihilation operator $d(z)$ is the modes into which photons are
scattered. 
A similar expression for the mode $b$ is obtained by replacing $a$ with $b$.

The observable used for the output detection schemes
in both Refs.~\cite{gilbert06,rubin07},
namely, $A = |N,0\rangle \langle 0,N| + |0,N\rangle \langle N,0|$,
is equivalent to the parity measurement
for the NOON state \cite{lee02}.
In addition, if we now only consider 
the measurement performed in the
{\em $N$-photon subspace} of the output state, 
we can ignore the
scattering term of the above transformation.

Following Ref.~\cite{gilbert06},  
we assume that
the losses are present only in 
the one of the two arms of the interferometer
and set $e^{-K_a L_a}=1$ and 
$e^{-K_b L_b}\equiv\ld$.
The associated operation of the lossy Mach-Zehnder
interferometer then can be
expression as 
\be \lx \bx a_{out} \\ b_{out} \ex \rx
&=&
\frac{1}{2} \lx \bx 1+\ld e^{i \vp} & -i(1-\ld e^{i \vp}) \\
i(1-\ld e^{i \vp}) & 1+\ld e^{i \vp} \ex\rx \lx \bx a_{in} \\
b_{in} \ex \rx 
, \label{ud} 
\ee 
which is non-unitary unless
$\ld = 1$. 
In the angular momentum representation, this
transformation can be rephrased as 
$L(\vp) = e^{iJ_x {\pi
\over 2}} \Lambda e^{-iJ_x {\pi \over 2}} e^{iJ_x {\pi \over 2}}
e^{-iJ_z {\vp}} e^{-iJ_x {\pi \over 2}} 
= e^{iJ_x {\pi
\over 2}} \Lambda e^{-iJ_x {\pi \over 2}} e^{-iJ_y {\vp}}$,
where $\Lambda $ is a matrix representing the effect of path
absorption. 
Then we get 
\be
L^{\dagger}P_N L &=& e^{iJ_y {\vp}} e^{iJ_x {\pi \over 2}} \Lambda
e^{-iJ_x {\pi \over 2}} P_N e^{iJ_x {\pi \over 2}} \Lambda
e^{-iJ_x {\pi \over 2}} e^{-iJ_y {\vp}} \nn 
& = & \ld ^N e^{iJ_y
{\vp}} P_N e^{-iJ_y {\vp}} \equiv \mathcal{Y}_1. 
\ee 
with $P_N=P \otimes \sum_{m=-j}^{m=j} |j,m\rangle \langle j, m|$ 
denoting the
$N$-photon projected parity operator.
That is to say,
one needs to detect all the $N$ photons,
even though that probability 
decreases exponentially with $N$.

Similarly, we find 
\be 
L^{\dagger}P_N^2 L &=& L^{\dagger} L =
e^{iJ_y {\vp}} e^{iJ_x {\pi \over 2}} \Lambda^2 e^{-iJ_x {\pi
\over 2}} e^{-iJ_y {\vp}} \nn 
&=& e^{iJ_x {\pi \over 2}} \Lambda^2
e^{-iJ_x {\pi \over 2}} \equiv \mathcal{Y}_2, 
\ee 
where the commutability of
$\mathcal{Y}_2$ and $e^{-iJ_y {\vp}}$ is applied, 
which can be
simply proved in the spinor representation. 
\begin{figure}[t!]
\begin{minipage}{0.3\textwidth}
\centerline{\epsfxsize 70mm \epsffile{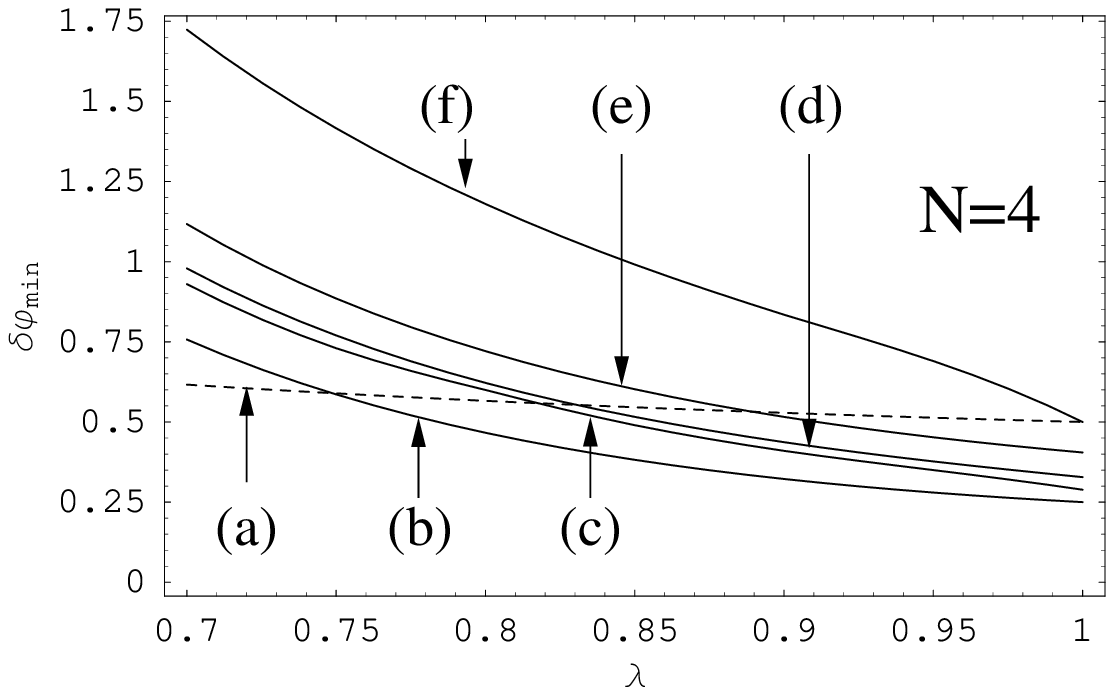}}
\end{minipage}
\begin{minipage}{0.3\textwidth}
\centerline{\epsfxsize 70mm \epsffile{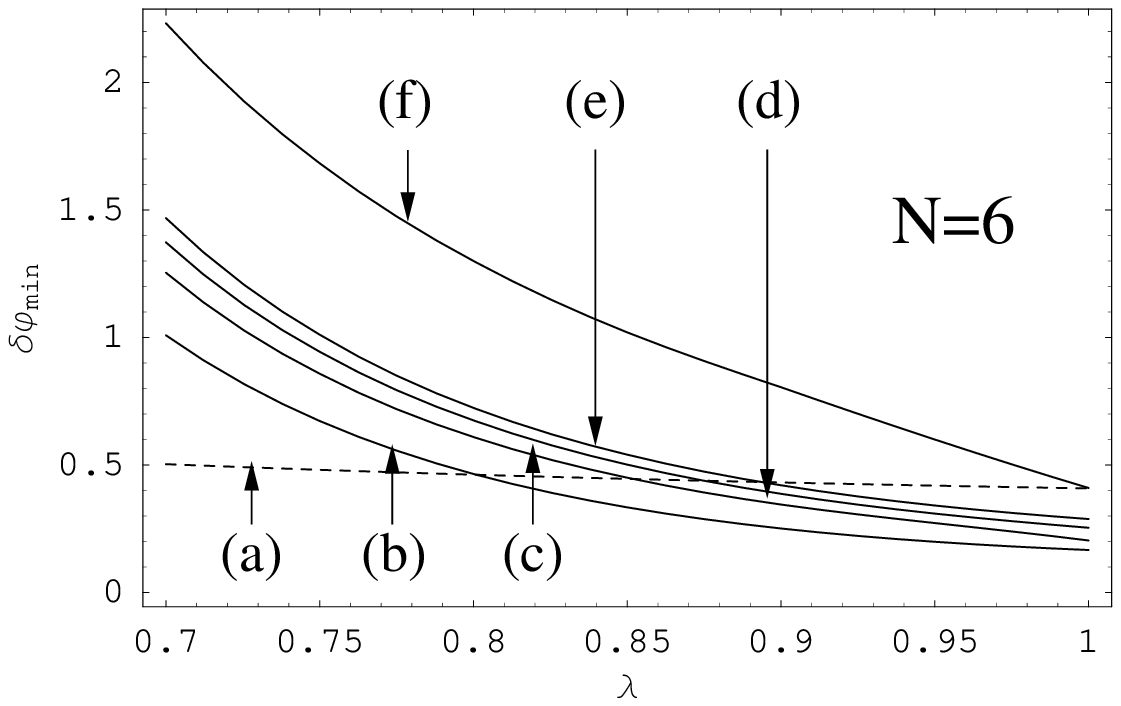}}
\end{minipage}
\caption{The minimum phase sensitivity, $\delta \vp_{\text {min}}$, for the
various entangled states as a function 
of  $\ld$, the transmission coefficient.
The upper and lower figures are for $N=4$ and $N=6$, respectively. 
The dotted line (a) represents that 
of the uncorrelated input state \cite{gilbert06}.
The solid lines represent (b) the NOON state,
(c) the dual Fock state, (d) the intelligent ($\eta=10$) state, 
(e) the Yurke, and (f) the intelligent ($\eta=1$) state, respectively.}
\end{figure}
Now, for a
general input state, $\lg {\text {in}}=\sum_{m=-j}^j c_m \lg {j,m}$, 
we obtain
$
\langle P_N \rangle_{\text{out}}
= \langle  \mathcal{Y}_1 \rangle_{\text{in}}
= (-1)^j \ld^{2j}
\sum_{m,n} c^*_m c_n (-1)^m d^j_{mn}(2\vp)
$, and
$
\langle P_N^2 \rangle_{\text{out}}
= \langle  \mathcal{Y}_2 \rangle_{\text{in}}
=
(1/2) \sum_{m,n} c^*_m c_n [Q_{mn}+Q_{nm}](\ld)
$.
Here, the
polynomial $Q_{mn}(\ld)$ is defined as the matrix element 
$\langle j,m | \mathcal{Y}_2 |j,n \rangle$ such that 
\be Q_{mn}(\ld) &=& { i^{-m-n}
\over (j-n+1)_{j+n}}\frac{(2 j)! \sqrt{(j+n)!} }{\sqrt{(j-m)!
(j+m)!(j-n)!}}\nn && \times
\left(\frac{x^2-1}{4}\right)^{j}\left(\frac{1+x}{1-x}\right)^{j+n-m}
\nn && \ts \!\ P_{j+n}^{(-2 j-1,m-n)} \left(1-\frac{8
x}{(x+1)^2}\right), 
\ee 
where $x \equiv \ld^2$ and $p_q \equiv
\Gamma(p+q)/\Gamma(p)$.

We now compare the phase sensitivity for different
entangled states in the presence of photon loss. 
The plots depicted in FIG.~2 show the reduced phase sensitivity 
due to the photon loss, in this case as a function 
of $\lambda$ (the transmission coefficient).
All the commonly proposed entangled
states are compared to the lossy-environment
shot-noise limit.
Among the entangled states,
the best possible 
phase sensitivity can be achieved by the NOON state,
and it gets worse in the following order:
the dual Fock state, the $\eta =10$ intelligent state,
the Yurke state,
and then the $\eta =1$ intelligent state.
Within the restricted parity measurement scheme
the NOON states show the best performance for phase detection and can
still beat the shot-noise limit if the transmittance of interferometer
is not too small and the photon number is not too large. 
We see that beating the shot-noise limit 
(dotted line, represented by the uncorrelated input state)
requires less attenuation as the number of photons increases.
For example, the lowest solid line (representing the NOON states) requires
75\% transmission for $N=4$ and 80\% for $N=6$.

To summarize, we showed that
the utilization of the parity measurement in  
sub-shot-noise interferometry
is applicable to a wide range of quantum entangled input states,
so far known entangled states of light.
Comparison of the performance of the various
quantum states then can be made within such
a unified output measurement scheme.
Furthermore,
it may lead to a great reduction of the efforts
in precise quantum state preparation as well as
in various optimization strategies involving
quantum state engineering for 
the sub-shot-noise interferometry \cite{gao08}.

The authors wish to thank J.P.\ Dowling for stimulating discussions, 
and would like to acknowledge support from 
US AFRL, ARO, IARPA, and DARPA.


\begin{thebibliography}{9}
\bibitem{bs}
For the sake of simplicity, $\rm BS_-$ is chosen
to be conjugate of $\rm BS_+$ such that, if all the photons enter 
the upper port, they will also come out at the upper port
in the absence of the relative
phase shift.
\end{thebibliography}
\end{document}